\documentclass{webofc}

\usepackage[varg]{txfonts}   
\usepackage{hyperref}
\usepackage{wrapfig}
\usepackage{url}

\hypersetup{colorlinks=true,citecolor=blue,urlcolor=blue,linkcolor=blue}

\newcommand{\imprfsc}{f_{G\chi}}

\begin{document}

\title{Finite-volume analysis and 
universal scaling signatures near the chiral phase transition in (2+1)-flavor QCD}

\author{\firstname{Sabarnya}  \lastname{Mitra}\inst{1}\fnsep\thanks{\email{smitra@physik.uni-bielefeld.de}} \and
        \firstname{Frithjof}\,  \lastname{Karsch}\inst{1}
}

\institute{Fakult\"at f\"ur Physik, Universit\"at Bielefeld, D-33615 Bielefeld,
Germany}

\abstract{
For quantifying the universal properties of the chiral phase 
transition in QCD through numerical calculations on a discrete space-time
lattice, one needs to perform controlled extrapolations to the continuum and
infinite-volume limits followed by an extrapolation to the limit of 
massless light quarks. We discuss here, the results on the latter two
limits at still finite lattice spacings. We use here for chiral symmetry breaking,
an improved order parameter free of additive and multiplicative divergences and we 
analyse its volume and quark mass dependence. Comparing to the expected 
universal behavior in the chiral limit, we quantify deviations from the universal finite-size
scaling behavior as function of the light to strange quark mass ratio.
}

\maketitle

\newcommand{\muB}{\mu_B}

\newcommand{\barechiralcondinlattice}{\left \langle \Bar{\psi}\,\psi \right \rangle}
\newcommand{\barechiralcond}{M_\ell}
\newcommand{\barechiralsusc}{\chi_\ell}
\newcommand{\barechiralcondreg}{M_{\ell, \text{reg}}}
\newcommand{\barechiralsuscreg}{\chi_{\ell, \text{reg}}}
\newcommand{\impchiralcondreg}{M_{\text{reg}}}
\newcommand{\condsclfunc}{f_G}
\newcommand{\suscsclfunc}{f_{\chi}}
\newcommand{\imprsclfunc}{f_{G\chi}}

\section{Introduction}
\label{intro}

In order to quantify universal (e.g. the critical exponents) and non-universal 
(e.g. the phase transition temperature) properties
of the chiral phase transition  
in Quantum Chromodynamics (QCD), a carefully controlled extrapolation 
to the
(i) continuum, (ii) infinite-volume and  (iii) chiral limits is needed.
This has been pursued in the first determination of the
chiral phase transition temperature in (2+1)-flavor QCD with smaller than 
physical light quark masses and a physical value for the strange quark
mass \cite{HotQCD:2019xnw}.
\begin{wrapfigure}{r}{0.5\textwidth}
	\vspace{-0.6cm}
\begin{center}
  \includegraphics[width=.45\textwidth]{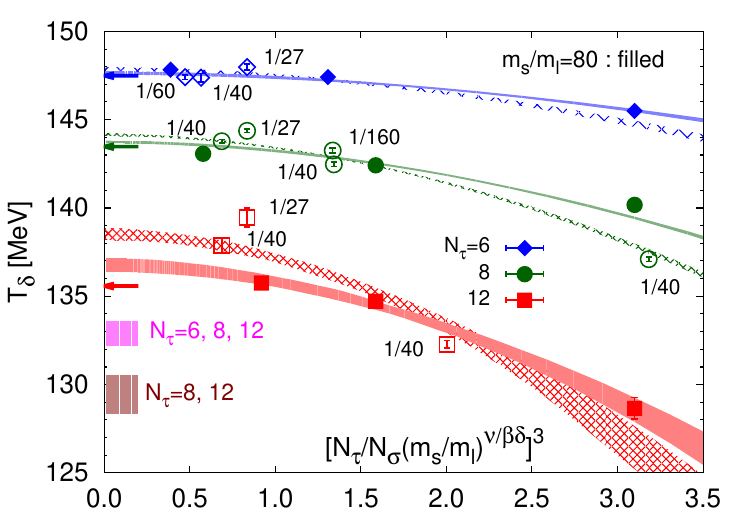}  
	\vspace{-0.6cm}
\end{center}
  \caption{Volume and quark mass extrapolation of pseudo-critical
	temperatures in (2+1)-flavor QCD with HISQ action \big(\,taken from \cite{HotQCD:2019xnw}\,\big).
}
\vspace{-0.9cm}
  \label{fig:pseudoTc}
\end{wrapfigure}
Results from an analysis of pseudo-critical temperatures in (2+1)-flavor QCD
obtained with the Highly Improved staggered quark (HISQ) action are shown in Fig.~\ref{fig:pseudoTc}. 
This analysis made use of known finite-volume scaling functions of the 3-$d$, $O(4)$
universality class \cite{Engels:2014bra} to perform joint infinite-volume and continuum limit 
extrapolations of pseudo-critical temperatures in calculations with light ($m_\ell$) to 
strange ($m_s$) 
quark mass ratios $H=m_\ell/m_s\ge 1/160$. In order to allow for a direct 
determination of universal critical exponents and thus derive
the underlying universality class for the chiral phase transition, one needs to
extend these calculations to smaller $H$ and perform a more systematic analysis
of finite-volume corrections that allows to identify the unique universal finite-volume contributions to physical observables. We present here first results for
finite-volume scaling corrections to the chiral order parameter on lattices with fixed temporal extent, $N_\tau$\,=\,$8$, varying the spatial extent $N_\sigma$. We compare
our results with expected universal critical behavior in the 3-$d$, $O(2)$ universality class and quantify the magnitude of sub-leading corrections.

\section{Volume dependence of the order parameter} 
\label{sec:Vol dep of ord param}

 The starting point of this work encompasses the multiplicatively renormalised versions of the $2$-flavor light quark chiral condensate, $\barechiralcond=m_s\,\barechiralcondinlattice/f_K^{\,4}$,
and the total chiral susceptibility $\barechiralsusc=m_s\,\partial M_\ell/\partial m_\ell$.
Here $f_K$ denotes the kaon decay constant. 
To attain a well-defined improved order parameter $M$ in the continuum limit,
which in the chiral limit also has a straightforward interpretation in terms of universal 
scaling functions,
we use the so-called subtracted order parameter \cite{Unger:2010wcq,Ding:2024sux}, 
where additive ultraviolet divergent contributions to $M_\ell$ are eliminated, 

 \begin{equation}
     M(T,H,L) = \barechiralcond \,(T,H,L) - H\,\barechiralsusc\,(T,H,L)\; .
     \label{eq:M}
 \end{equation}
Here $L$\,=\,$N_\sigma/N_\tau$ denotes the aspect ratio of a $N_\sigma^3$ $\cdot$ $N_\tau$ lattice.

At non-vanishing lattice spacing `$a$', the temperature and the  volume are given by $T=(N_\tau a)^{-1}$
and $V=(N_\sigma a)^3$.
We also note that contributions linear in $H$, present otherwise in the regular and singular parts of $M_\ell$, are explicitly cancelled in the definition of the 
order parameter $M$ given in Eq.\eqref{eq:M}, consequently reducing the non-universal regular part.  In the vicinity of the chiral critical
\begin{wrapfigure}{r}{0.5\textwidth}
        \vspace{-0.6cm}
\begin{center}
  \includegraphics[width=.45\textwidth]{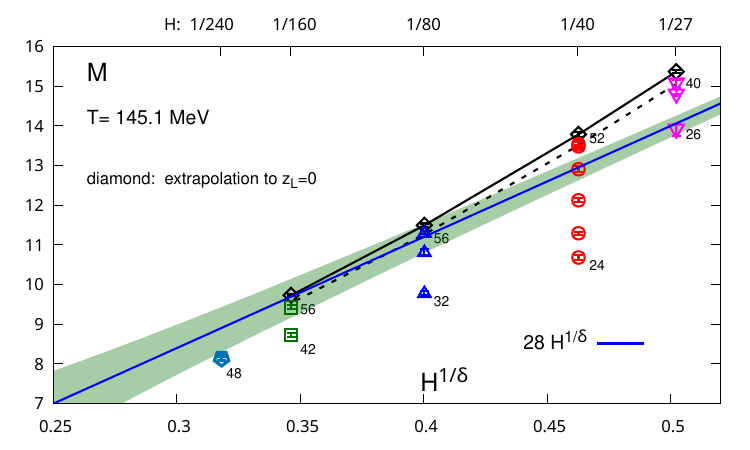}
        \vspace{-0.6cm}
\end{center}
  \caption{Results of order parameter $M$ as function of $H^{1/\delta}$ at $T=145.1$ MeV obtained on several $N_\sigma^3 \cdot 8$ lattices, with varying $N_\sigma$\,. 
  Here, $\delta=4.7798$ and $H=m_\ell/m_s$\,. 
  Black diamond points show extrapolated infinite-volume limit ($z_L=0$) values for different $H$, or light quark masses $m_\ell$\,. The green band shows a 0.5~MeV error band 
  on our new preliminary estimate for $T_c$ on $N_\tau=8$ lattices and the new, preliminary slope 
  of the blue line $M=m_0$\,$H^{1/\delta}$ is taken to be $m_0=28$, which is $\sim 10\%$
  smaller than the value for $m_0=\left(1-1/\delta\right)h_0^{-1/\delta}$\,, obtained in
  \cite{Karsch:2023pga}. 
}
  \label{fig:Mvol}
\end{wrapfigure}
point $(t,h,l)$\,=\,$(0,0,0)$, one has,
\begin{equation}
    M(T,H,L) = h^{1/\delta} \imprfsc (z,z_L) + M_{sub} \; ,
    \label{eq:M-scaling}
\end{equation}
where $\imprfsc$\,=\,$f_G$\,-\,$f_\chi$ with $f_G,\, f_\chi$ the respective scaling functions of $M_\ell$ and $\chi_\ell$, and $M_{sub}$ refers to sub-dominant contributions arising from corrections-to-scaling as well as regular terms.
 The scaling variables $z,z_L$  are as follows, 
 \begin{equation}
     z = t\,h^{-1/\beta\delta}, \hspace{.2cm} z_L=l\,h^{-\nu_c}\; , 
     \label{eq:scaling variable}
 \end{equation}
where $t=\left(T/T_c-1\right)/t_0$   
with chiral phase transition temperature $T_c$\,,  $h = H/h_0\,,\, l$\,\,=\,\,$l_0/L$, along with the critical exponents $\beta,\,\delta$ and $\nu$ 
of the associated $3$-$d$ universality class with $\nu_c$\,\,=\,\,$\nu/\beta\delta$\,\,=\,\,$(1+1/\delta)/3$. Further notations and detailed discussions are outlined in \cite{Karsch:2023pga,Mitra:2024mke,Mitra:2025ofj}. 
The dominant universal term of Eq.\,\eqref{eq:M-scaling}
can be directly exploited to derive the underlying universality class since no conclusive first order evidence has been found yet here in (2+1)-flavor QCD on lattice \cite{Cuteri:2021ikv}. 
However, as mentioned before one requires for this purpose, a detailed analysis at small quark mass and on large lattices to control the chiral and infinite-volume limits. This systematically diminishes contributions from $M_{sub}$, given in Eq.\eqref{eq:M-scaling}, as well as finite-volume corrections to the underlying universal behavior, thereby augmenting the latter. We illustrate some preliminary results in the next section.

\section{Universal finite-volume effects}
\label{sec: Univ fin vol effects}

\begin{figure*}
\centering
\includegraphics[scale=.48]{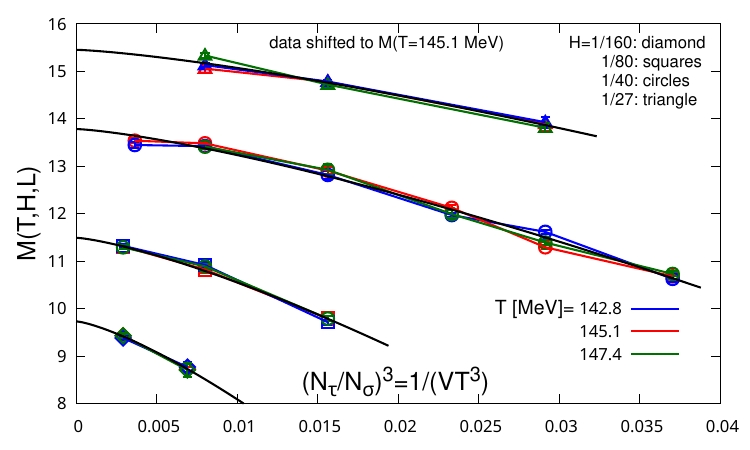}
\includegraphics[scale=.48]{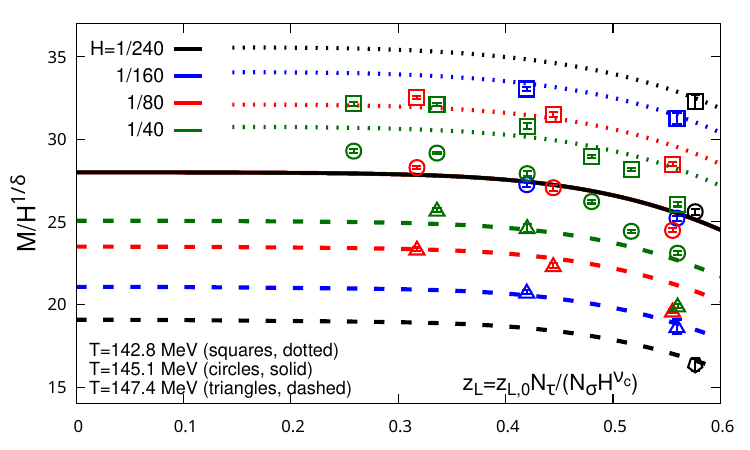}
\caption{{\it Left:} Finite-volume dependence of the order parameter $M$. Shown are 
results for $M$ versus the inverse volume normalized by $T^3$  
obtained on lattice with temporal extent $N_\tau=8$ at four values of the quark mass ratio $H$ and three temperature values in the vicinity of $T_c\simeq 145$~MeV. Data obtained at $T=142.8$~MeV
and $T=147.4$~MeV have been shifted by single, constant values.
{\it Right:}
The scaled order parameter $M/H^{1/\delta}$ shown as a function of the finite-volume 
scaling variable $z_L$. Lines show the scaling function $f_{G\chi}(z,z_L)$ for 
three temperatures using preliminary results for the non-universal scale parameters,  
$T_c=145.1$~MeV, $z_0=h_0^{1/\beta\delta}\big/\,t_0=1.52$ and $z_{L,0}=l_0\,h_0^{\nu_c}=0.38$. 
}
\vspace{-0.2cm}
\label{fig:fin-vol-dep-of-M}   
\end{figure*}
In this work, we have used tree-level Symanzik-improved gauge and HISQ actions 
implemented in \textit{SIMULATeQCD} \cite{Mazur:2021the,Mazur:2023sim} for 
numerical calculations.
While keeping the strange quark mass tuned to its physical value, the light quark masses are varied in a range corresponding to pion masses $m_\pi \in$ [45:140] MeV.
We substantially extended existing data sets for $H \in$ [1/160\,:\,1/27] 
and 
included new data sets with $H=1/240$ for $T \in$ [143:147] MeV to probe the universal behavior. 
Results for $M$ obtained for various $H$ values and lattice sizes $N_\sigma$
are shown in Fig.\,\ref{fig:Mvol}. For each vale of $H$ the smallest and largest value
of $N_\sigma$ is given next to the data points. Results are shown for $T=145.1$~MeV,
which is our new preliminary estimate for $T_c$ on $N_\tau=8$ lattices. 
We will compare the numerical results with the finite-volume scaling function
$f_{G\chi}(z,z_L)$
for the to 3-$d$ $O(2)$ universality class, which is appropriate for calculations with
staggered fermions at non-vanishing `$a$'. In this scaling function, finite-size
corrections are treated in terms of a Taylor series 
in $z$ and $z_L$ \cite{Karsch:2023pga} : 

\begin{equation}
    \imprfsc(z,z_L) = \imprfsc (z,0) + \sum_{m=m_l}^{m_u} \left(1-\frac{1}{\delta}+\frac{m\nu}{\beta\delta}\right)\,a_{0m}\,z_L^{m} \,+\, \sum_{n=1}^{n_u} \sum_{m=m_l}^{m_u} \left(1-\frac{1}{\delta}+\frac{n+m\nu}{\beta\delta}\right)\, a_{nm}\,z^{n} z_L^{m}\; .
    \label{eq:fGchi finite vol}
\end{equation}
 While the second term of Eq.\eqref{eq:fGchi finite vol} projects the finite-volume corrections at $T$\,=\,$T_c$, the first term depicts the infinite-volume limit ($z_L$\,=\,$0$) of $\imprfsc$ at different $T$ which, at $T_c$ is $\imprfsc(0,0)=\left(1-1/\delta\right)$. For fixed $T$ and $H$, we performed
 infinite-volume extrapolations using an ansatz
 inspired by Eq.\eqref{eq:fGchi finite vol}, which however does not make any explicit use of universal properties :  

 \begin{equation}
  f\left(T,H,L\right)=a_0\left(T,H\right)\,+\,a_3\left(T,H\right)/\,L^3\,+\,a_4\left(T,H\right)/\,L^4   \; .  
 \end{equation}

 As apparent from Fig.\,\ref{fig:Mvol}, for $H>1/80$, the $z_L$\,=\,$0$ data points
 show significant deviations from an expected scaling behavior $M/H^{1/\delta} = h_0^{-1/\delta}$
 (solid line) valid at $T_c$. The dashed line in Fig.\,\ref{fig:Mvol} shows a 2\% deviation from
 the infinite-volume extrapolated values. It is evident that the aspect ratio needed to reach
 such an accuracy increases from $N_\sigma/N_\tau\simeq 5$ at $H=1/27$ to $N_\sigma/N_\tau\simeq 7$ at $H=1/160$. 
 One finds similar likewise implications for finite-volume effects in the right Fig.\,\ref{fig:fin-vol-dep-of-M} plot, 
 which shows the rescaled order parameter $M/H^{1/\delta}$ versus the finite-size scaling variable 
 $z_L$ for three different values of the temperature. Unlike Fig.\,\ref{fig:Mvol}, this figure 
 makes explicit use of universal parameters in the 3-$d$, $O(2)$ universality class.
 It is evident that $H \ge 1/80$ data points deviate from the respective scaling curves. This is more severe for higher $H=1/40$ with noticeable discrepancies even for large lattice volumes ($z_L \sim 0.3$).  
 On the other hand, $H$\,=\,$1/160,1/240$ points show commendable agreement with expected 
 universal finite-volume scaling even for $z_L \sim 0.6$, corresponding to smaller lattice volumes. Surely, additional $1/240$ data
 sets for larger volumes are still needed to arrive at firm conclusions. 
 We also note that in a small temperature range around $T_c$ {\it i.e.} $142.8~{\rm MeV}\le T\le 147.4~{\rm MeV}$, finite-volume effects are temperature independent to a large degree.
 This is observed in the left Fig.\,\ref{fig:fin-vol-dep-of-M} plot, where the data points for $T=142.8$ and $147.4$ MeV, with  $H \in $ [$1/160$\,:\,$1/27$]\,, $N_\sigma \in$ [$24$\,:\,$56$] 
 when shifted to $T=145.1$ MeV, shows promising coincidence.

\section{Summary and Outlook}
\label{sec: Summ and outlook}

We have demonstrated here that a systematic finite-volume analysis with controlled infinite-volume extrapolation for smaller quark mass values close to $T_c$ indeed exhibit universal signatures with
subdued finite-volume effects and sub-leading contributions. 
This facilitates a more precise estimate of $T_c$ thereby providing a more stringent bound on QCD critical point \cite{Karsch:2019mbv}, apart from quantitatively estimating the scaling regime and determining the underlying universality class. Nevertheless, more data close to the chiral critical point are still to be added besides, planning future simulations on $N_\tau$\,=\,$12$ lattices. This is to obtain good enough precision for differentiating $U(2) \times U(2)$ and $O(4)$ universality classes in the continuum limit, which we hope would also potentially resolve the long-standing controversially discussed fate of $U(1)_A$ symmetry at the chiral phase transition temperature \cite{Pisarski:1983ms,Pelissetto:2013hqa}.

\section*{Acknowledgements}

This work is supported by the Deutsche Forschungsgemeinschaft (DFG, German Research Foundation) Project No. 315477589-TRR 211. We thank the Bielefeld HPC.NRW and the NHR Center PC2 Paderborn teams for the computing times in the respective GPU clusters.

\end{document}